\documentclass[aps,preprint,prd,epsfig]{revtex4}
\pdfoutput=1
\usepackage{amssymb}
\usepackage{amsmath}
\usepackage{graphicx}
\usepackage{color}
\usepackage{fancyhdr}
\newcommand{\be}{\begin{equation}}
\newcommand{\ee}{\end{equation}}
\newcommand{\bea}{\begin{eqnarray}}
\newcommand{\eea}{\end{eqnarray}}
\newcommand{\bes}{\begin{subequations}}
\newcommand{\ees}{\end{subequations}}

\newcommand{\bc}{\begin{center}}
\newcommand{\ec}{\end{center}}
\usepackage{amsmath, amssymb} 
\usepackage{slashed}
\usepackage{natbib}

\begin{document}

\title{Confronting  the inverse seesaw  mechanism with the recent muon g-2 result}

\author{Jo\~ao Paulo Pinheiro$^{a}$, C. A. de S. Pires$^{b}$, Farinaldo S. Queiroz$^{c,d,e}$, Yoxara S. Villamizar$^{c,d}$}
\affiliation{{$^a$ Departament de Fisica Quantica i Astrofisica and Institut de Ciencies del Cosmos, Universitat de Barcelona, Diagonal 647, E-08028 Barcelona, Spain},\\
{$^b$ Departamento de Física, Universidade Federal da Paraíba, Caixa Postal 5008, 58051-970, Jo\~ao Pessoa, PB, Brazil}, \\
{$^c$Departamento de F\'isica, Universidade Federal do Rio Grande do Norte, 59078-970, Natal, RN, Brasil}, \\
{$^d$International Institute of Physics, Universidade Federal do Rio Grande do Norte, Campus Universitario, Lagoa Nova, Natal-RN 59078-970, Brazil}, \\
{$^e$ Millennium Institute for Subatomic Physics at High-Energy Frontier (SAPHIR), Fern\'andez Concha 700, Santiago, Chile}
}

\date{\today}

\begin{abstract}
Since the heavy neutrinos of the inverse seesaw mechanism mix largely  with the standard ones, the charged currents formed with them and the muons  have the potential of generating robust and positive contribution to the anomalous magnetic moment of the muon. Ho\-we\-ver, bounds from the non-unitary in the leptonic mixing matrix may restrict so severely the parameters of the mechanism that, depending on the framework under which the mechanism is implemented, may render it unable to explain  the recent muon g-2 result. In this paper we show that this happens when we implement the mechanism into the standard model and into two versions of the 3-3-1 models. 

\end{abstract}

\maketitle

\section{Introduction}

The recent measurement of   $a_\mu \equiv \frac{g-2}{2} $ by the E989 experiment at Fermilab \cite{Muong-2:2021ojo}, in agreement with the previous BNL E821 result \cite{Muong-2:2006rrc}, yields a $4.2\sigma$ discrepancy from the standard model (SM) $$\Delta a_\mu = a_{\mu}^{\mbox{exp}} - a_{\mu}^{\mbox{SM}} = (251 \pm 59)\times 10^{-11}.$$  It represents a hint of physics beyond standard model (BSM) \cite{Keshavarzi:2021eqa}. The easiest way to generate new contributions to the g-2 of the muon is by means of new interactions involving the leptonic sector (See Ref.\cite{Lindner:2016bgg} for an extended review).  Furthermore,  the SM cannot address neutrino masses. Therefore, it is appealing to investigate models that can accommodate both problems. 

There are several ways to explain neutrino masses, but in this work, we will focus on the inverse seesaw (ISS) mechanism\cite{Mohapatra:1986bd, Mohapatra:1986aw, Dias:2011sq}. It is a genuine TeV scale seesaw mechanism whose signature, heavy neutrinos $N$'s, is supposed to manifest at TeV scale. This interesting aspect allows us to confront the mechanism with existing TeV scale probes. In light of the recent measurement conducted by the Fermilab g-2 experiment, one might wonder whether one can successfully explain g-2 with the ISS.

The minimum SM extension capable of embedding the ISS has been addressed previously to analyze the corresponding g-2 contributions for sterile fermion states \cite{abada2014effect}. We consider the contributions for the g-2  of the muon due exclusively to the heavy neutrinos of the ISS mechanism. To do this, we consider models that do not accommodate the recent muon g-2 result (SM and some versions of the 3-3-1 model) and implement the ISS mechanism in them economically, where only the new neutral fermion singlets are added to the models in question. We then try to obtain a region of parameter space that could accommodate the recent measurement of g-2  of the muon and agree with some bound over the mechanism.

Our work is organized in the following way: In Sec. II, we present the main ingredients of the mechanism and develop our approach. In Sec. III, we implement the mechanism into the standard model, the minimal 3-3-1 model and the 3-3-1 model with right-handed neutrinos and explicit the terms that give the main contributions to the g-2  of the muon due to the heavy neutrinos. In sec. IV, we present our numerical results. We finalize presenting our conclusions in Sec. V.

\section{The main ingredients of the  mechanism}
\subsection{Framework and nonunitarity costraint}
The mechanism involves nine neutral fermions with specific chirality $\nu_L =(\nu_{e_L}\,,\, \nu_{\mu_L}\,,\,\nu_{\tau_L})$, $S_R=(S_{e_R}\,,\,S_{\mu_R}\,,\,S_{\tau_R})$ and $S_L=(S_{e_L}\,,\,S_{\mu_L}\,,\,S_{\tau_L})$ composing the following mass terms
\begin{eqnarray}
{\cal L}=-\bar{\nu_L}M_D S_R - \Bar{S_L}MS_R-\dfrac{1}{2}\Bar{S_L}\mu S^C_L +H.c. ,
\label{generalmassterms}
\end{eqnarray}
where $M_D$, $M$ and $\mu$ are generic $3\times 3$ mass matrices. We rearrange things so that in the basis $\nu=(\nu_L, S^{ C}_L, S_L)$ we may write
\begin{eqnarray}
{\cal L}=-\frac{1}{2}\nu^C M_ \nu  \nu +H.c. ,
\end{eqnarray}
where,
\begin{equation}
    M_\nu=
    \begin{pmatrix}
    0&M_D^T&0\\
    M_D&0&M^T\\
    0&M&\mu
    \end{pmatrix}.
    \label{mainmatrix}
\end{equation}

In recognizing, 
\begin{equation}
{\cal M_D}_{6\times 3}=
\begin{pmatrix}
{M_D}_{3\times 3}\\
0_{3\times 3}
\end{pmatrix},\;\;\;\;
{\cal M_R}_{6\times 6}=
\begin{pmatrix}
0_{3\times 3}&{M^T}_{3\times 3}\\
M_{3\times 3}&\mu_{3\times 3}
\end{pmatrix},
\label{M66}
\end{equation}
we write
\begin{equation}
{ M_\nu}=
\begin{pmatrix}
0_{3\times 3}&{\cal M_D}^T_{3\times 6}\\
{\cal M_D}_{6\times 3}&{\cal M_R}_{6\times 6}
\end{pmatrix}.
\end{equation}

This mass matrix may be diagonalized by a mixing matrix  $W$ given by\cite{Schechter:1981cv,Hettmansperger:2011bt}
\begin{equation}
W\approx
\begin{pmatrix}
1-\dfrac{1}{2}({\cal M_D})^\dagger [{\cal M_R}({\cal M_R})^\dagger]^{-1}{\cal M_D}&({\cal M_D})^\dagger[{\cal M_R}^\dagger]^{-1}\\
-[{\cal M_R}]^{-1}{\cal M_D}&1-\dfrac{1}{2}({\cal M_R})^{-1} {\cal M_D}({\cal M_D})^\dagger[{\cal M_R}^\dagger]^{-1}
\end{pmatrix},
\end{equation}
such that

\begin{equation}
    W^T{ M_\nu}W=\begin{pmatrix}
    {m_{light}}_{3\times3}&0_{3\times 6}\\
    0_{6\times 3}&{m_{heavy}}_{6\times6}
    \end{pmatrix},
\end{equation}
where we have 
$m_{light}=-{\cal M_D}^T{\cal M_R}^{-1}{\cal M_D},\;\;\;m_{heavy}={\cal M_R}$.
In terms of the original parameters, we have,
\begin{equation}
m_{light}= -M_D^T M^{-1}\mu {M}^{-1} M_D.
\label{ISSmass}
\end{equation}
Observe that the diagonalization of  $m_{light}$ must lead to the eigenvalues $m=(m_1\,,\,m_2\,,\,m_3)$, which we assume are associated with the following  eigenvectors  $n_L=( n_{1_L}\,,\,n_{2_L}\,,\,n_{3_L})$. 

Now comes the essence of the mechanism. For $m_D$ belonging to the electroweak  and $M$ to the TeV scale, we just need that lepton number be explicitly violated at sub-keV  scale ($\mu \sim 0.1$keV) in order to have $m$ at sub-eV scale. Of course, we are assuming that $n_L$ are the standard neutrinos.

Concerning the heavy neutrinos, we assume that  $m_{heavy}$ is diagonalized by $U_R$, and that the eigenvectors are given by  $N_L=N_{i_L}$, with $i=1...6$, whose masses lie around $M$. Moreover, the mixing among the neutrinos allows us to write the standard flavor neutrinos as the following  combinations of the physical neutrinos\cite{ Dev:2009aw}
\begin{eqnarray}
\label{mass_s}
\nu_{\alpha L}=[U_{PMNS}\left(1-\dfrac{1}{2}(F^\dagger F)\right)]_{\alpha i}n_{i L }+[{\cal M_D}^\dagger({\cal M_R}^\dagger)^{-1}U_R]_{\alpha k }N_{k L},
\end{eqnarray}
where $F=M_DM^{-1}$. In view of the structure of this mixing,
all deviation from unitarity is  determined by the Hermitian matrix, 
\begin{equation}
    \eta=\frac{1}{2}F^{\dagger}F.
\label{etaparameter}
\end{equation}
We point out that it does not depend on the parametrization
of the PMNS matrix. This $\eta$ matrix is known as the nonunitarity parameter and have a crucial role in this paper. Current bounds on nonunitarity effects gives \cite{Fernandez-Martinez:2016lgt}

\begin{equation}
   \mid \eta_{\mbox{bound}} \mid < \begin{pmatrix}
    2.5\times 10^{-3}&2.4\times 10^{-5}&2.7\times 10^{-3}\\
    2.4\times 10^{-5}&4.0\times 10^{-4}&1.2\times 10^{-3}\\
    2.7\times 10^{-3}&1.2\times 10^{-3}&5.6\times 10^{-3}
    \end{pmatrix}.
    \label{nonunitarityconstraint}
\end{equation}

\subsection{Our approach}
In the usual approach, the matrices $M$ and $\mu$ are considered as diagonal ones. In this case  $M_D$ determines the texture of $m_{light}$. 

We take a different approach now. We assume that $M$ and $M_D$ are diagonal and degenerated: $M = diag(m_N,m_N,m_N)$ and $M_D=diag(m_{D},m_{D},m_{D})$. Then Eq. (\ref{ISSmass}) get
\begin{equation}
    m_{light}\approx \frac{m_D^2}{m_N^2} \mu,
 \label{mlight}   
\end{equation}
which leaves $\mu$  responsible for the texture of $m_{light}$. It happens that 
in our approach the parameter $\eta$, defined in Eq. (\ref{etaparameter}), takes the form

\begin{equation}
\eta=\begin{pmatrix}
\dfrac{m_{D}^2}{2m_N^2}&0&0\\
0&\dfrac{m_{D}^2}{2m_N^2}&0\\
0&0&\dfrac{m_{D}^2}{2m_N^2}
\end{pmatrix},
\label{etaSM}
\end{equation}
which, according to the bound in Eq. (\ref{nonunitarityconstraint}), yields,
\begin{eqnarray}
 \dfrac{m_D}{m_N}<0.04.
\end{eqnarray}
This is nice because the nonunitarity constraint is favoring $\mu$ below keV scale to yield standard neutrino masses at eV scale. 


After diagonalizing $m_{light}$ above, we obtain 
\begin{equation}
m_\nu = \dfrac{m_D^2}{m_N^2}U_{PMNS}^T\mu U_{PMNS},
\end{equation}
where $m_\nu = diag(m_1,m_2,m_3)$. Inverting these matrices, we get, 

$$\dfrac{m_N^2}{m_D^2} U_{PMNS}m_\nu U_{PMNS}^T = \mu . $$  
Once we know the absolute values of $m_1\,,\,m_2\,,\,m_3$ and having $U_{PMNS}$ from the experiments, we obtain the texture of $\mu$.



Now we focus on the pattern of $U_R$. For this we diagonalize $m_{heavy}$ given in Eq. (\ref{M66}) which in our approach:
\begin{equation} 
m_{heavy}=\begin{pmatrix}
0_{3\times 3}&m_N I_{3\times 3}\\
m_N I_{3\times 3}& \mu_{3 \times 3}
\end{pmatrix} =m_N \begin{pmatrix}
0_{3\times 3}& I_{3\times 3}\\
 I_{3\times 3}& \frac{\mu_{3 \times 3}}{m_N} 
\end{pmatrix}.
\end{equation}




For   $m_N>>\mid \mu_{ij}\mid$ the eigenvalues of $m_{heavy}$ are,
\begin{eqnarray}
 m_{N1,2,3} &=&m_N,\;\;m_{N4,5,6}  =-m_N,
\end{eqnarray}
which means that the six heavy neutrinos are practically mass degenerate with the  values lying around  $m_N$. In this case, the mixing matrix, $U_R$, that diagonalizes $m_{heavy}$ develops the following approximated pattern

\begin{equation} \small
U_R\approx
   \dfrac{1}{\sqrt{2}} \begin{pmatrix}
    0&0&-1&0&0&1\\
    0&-1&0&0&1&0\\
    -1&0&0&1&0&0\\
    0&  0&1&0&0&1\\
    0&1&0&0&1&0\\
    1&0&0& 1 & 0 &0\\
    \end{pmatrix} 
\end{equation}

In summary, in our approach, $\mu$ is responsible for the texture of the standard neutrino mass matrix, while the assumption  $m_N>>\mid \mu_{ij}\mid$ infers the  pattern of $U_R$. This method eases the assessment of the heavy neutrinos contributions to g-2 significantly consistently. With this method at hand, we can solidly investigate if the inverse seesaw mechanism may explain g-2 \cite{Muong-2:2021ojo}.

We remind the reader that we will adopt throughout the most recent result from the Muon g-2 Experiment at Fermilab, which combined with  previous Brookhaven National Laboratory E821 measurement reported $ a_{\mu}^{\mbox{exp}} = (g_\mu - 2)/2 = 16592061(41) \times 10^{- 11}$. The difference
$\Delta a_\mu= a_{\mu}^{\mbox{exp}} - a_{\mu}^{\mbox{SM}} = (251 \pm 59)\times 10^{- 11}$ has a significance of $4.2\sigma$\cite{Muong-2:2021ojo, Keshavarzi:2021eqa}. See Ref. \cite{Lindner:2016bgg} for an extensive explanation about how the anomaly can be weakened or strengthened.

The implementation of the ISS requires the introduction of new neutral fermions, $N_i$, to the original fermion content of the model in question. In extended gauge theories that feature the presence of new vector gauge bosons, such neutral fermions will appear in the charged currents involving exotic charged gauge bosons represented by $W^{\prime \pm}$. In a general way, we write down these interactions through the following terms

\begin{eqnarray}
 {\cal L}_{int}=g^{ij}_{v 1}W'^+_\mu{\bar N}_i\gamma^\mu l_j+g^{ij}_{a 1}W'^+_\mu{\bar N}_i\gamma^\mu \gamma_5 l_j + H.c.
 \label{newCC}
\end{eqnarray}
After some steps, the contribution of these new interactions to the g-2 of the muon is given by\cite{Lindner:2016bgg},
\begin{eqnarray}
 \Delta a_\mu(N,W') = -\dfrac{1}{8 \pi^2}\dfrac{m_\mu^2}{m_{W'}^2}\int^1_0 dx \sum_f \dfrac{\mid g^{ij}_{v 1} \mid^2 P^+_3(x)+\mid g^{ij}_{a 1} \mid^2 P^-_3(x)}{\epsilon_f^2\lambda^2(1-x)(1-\epsilon_f^{-2}x)+x}
\label{aN}
\end{eqnarray}
where

\begin{eqnarray}
 P_3^{\pm}=-2x^2(1+x\mp 2\epsilon_f) + \lambda^2x(1-x)(1\mp \epsilon_f)^2(x\pm \epsilon_f)
\end{eqnarray}
and $\epsilon_f \equiv \dfrac{m_{N_f}}{m_\mu}$, $\lambda \equiv \dfrac{m_\mu}{m_{W'}}$.

We apply this method to obtain the contribution to the g-2 stemming exclusively from the new ingredients of the ISS mechanism when implemented in models from starters that do not accommodate g-2.
\section{ISS contributions to the g-2  of the muon}
\subsection{Standard model}
To implement the ISS mechanism in a minimal way into the standard model, we need to add six new neutral fermions as singlets field, so the leptonic content becomes,

\begin{align}
l_{a_L}=\left(\begin{array}{c}
\nu_{a} \\ 
\ell_{a} 
\end{array}\right)_L,\,\,\, e_{a_R}\,,\,\nu_{e_R}\,,\,\nu_{\mu_R}\,,\,\nu_{\tau_R}\,,\, S_{e_L}\,,\,S_{\mu_L}\,,\,S_{\tau_L},
\end{align}
where $a=e\,,\,\mu\,,\,\tau$. The scalar sector is assumed to be composed uniquely by the standard Higgs $H$

With these new fields we can form the following new terms in addition to all standard ones,
\begin{equation}
    {\cal L}=  Y_D \bar l_L \tilde H \nu_R +\bar \nu_R M S_L + \frac{1}{2}\bar S^C_L \mu S_L + H.C., 
\end{equation}
where $\nu_R=(\nu_{e_R}\,,\,\nu_{\mu_R}\,,\,\nu_{\tau_R})^T$ and $S_L=(S_{e_L}\,,\,S_{\mu_L}\,,\,S_{\tau_L})^T$. When of the spontaneous breaking of the symmetry, where $\langle H \rangle=\frac{v}{\sqrt{2}}$, we obtain the following mass terms,
\begin{equation}
    {\cal L}=  \bar \nu_L M_D \nu_R +\bar \nu_R M S_L + \frac{1}{2}\bar S^C_L \mu S_L + H.C., 
    \label{ISSSM}
\end{equation}
where $M_D= \frac{Y_Dv}{\sqrt{2}}$ and $\nu_L=(\nu_{e_L}\,,\,\nu_{\mu_L}\,,\,\nu_{\tau_L})^T$. 

Observe that Eq. (\ref{ISSSM}) recovers Eq. (\ref{generalmassterms})  which means that we have the ISS mechanism. Thus, all the step done above is valid  here. 

Through mixing, Eq. (\ref{mass_s}), the ISS mechanism gives rise to the following Lagrangian,

\begin{eqnarray}
 {\cal L}^{CC}_{ISS}=-\dfrac{g}{2\sqrt{2}}{\bar \mu } (1-\gamma_5)\gamma^\mu [({\cal M_D}^\dagger)[{\cal M_R}^\dagger]^{-1}U_R]_{2k}N_{k}W^-_\mu + H.c. 
 \label{ISSCCSM}
\end{eqnarray}where $N_k$ are heavy pseudo-Dirac neutrinos. Notice that the product $({\cal M_D}^\dagger)[{\cal M_R}^\dagger]^{-1}U_R$ can be approximated by,

\begin{equation}
({\cal M_D}^\dagger)[{\cal M_R}^\dagger]^{-1}U_R\approx
    \dfrac{m_D}{\sqrt{2}m_N}\begin{pmatrix}
    0&0&1&0&0&1\\
    0&1&0&0&1&0\\
    1&0&0&1&0&0
    \end{pmatrix}
\end{equation}

It is clear that the six new pseudo-Dirac neutrinos  contribute to g-2. Then, from Eq. (\ref{ISSCCSM}) and Eq. (\ref{newCC}) we can easily identify $g_{v1}^{ij}$, and $g_{a1}^{ij}$  that appear in Eq. (\ref{aN}). 
We also set $\lambda=m_\mu/m_W$ and $\epsilon_f=\epsilon=m_N/m_\mu$. After all these considerations, we are ready to calculate the contribution from Pseudo-Dirac $N$'s to g-2  due to the  interactions in  Eq. (\ref{ISSCCSM}).




\subsection{Minimal 3-3-1 model}

This model is based on the ${SU(3)}_{{C}} \times {SU(3)}_{{L}} \times {U(1)}_{{X}}$ (3-3-1) gauge group\cite{Pisano:1992bxx,Frampton:1992wt}. In it, all  leptons of each family compose a triplet  in the following way, 
\begin{align}
f_{a_L}=\left(\begin{array}{c}
\nu_{a} \\ 
\ell_{a} \\
\left(\ell^{c}\right)_{a}
\end{array}\right)_L,
\end{align}
where $a=e\,,\,\mu \,,\, \tau$. Through this feature, we can understand the quantization pattern of electric charges \cite{deSousaPires:1998jc}. Moreover, the gauge anomalies are canceled only when three families are considered at once \cite{Foot:1992rh}. There are several additional interesting aspects \cite{Dias:2003iq,Pal:1994ba,Dias:2002hz,Alves:2011kc,Alves:2012yp}. The gauge sector of the model features the standard gauge bosons and five other ones as a new $Z^{\prime}$, two new single charged gauge bosons $W^{\prime \pm}$ and two doubly charged gauge bosons $U^{\pm \pm}$. Their respective neutral and charged currents are found in \cite{Cao:2016uur}. It has been shown in  \cite{deJesus:2020ngn} that all these interactions are not enough to accommodate the recent muon g-2 result.  Hence, it brings up whether the ISS mechanism can foot the bill while successfully generating small masses.  

For the implementation of the mechanism, we add the six neutral fermions $\nu_R$ and, $S_L$ as discussed previously. The minimal set of scalars necessary to break spontaneously the symmetry and generate masses for all massive particles, except neutrinos, are three triplets  $\eta'$, $\rho$, $\chi$ and a sextet of scalar $S$.  With this we form the terms
\begin{equation}
    {\cal L}\supset Y_D \bar f_L \eta' \nu_R +\bar \nu_R M S_L +\frac{1}{2}\bar S^C_L \mu S_L + H.C.
\end{equation}
When $\eta'$ develop vev, $\langle \eta' \rangle=\frac{v_\eta}{\sqrt{2}}$, we have the following mass terms,
\begin{equation}
    {\cal L} = \bar \nu_L M_D \nu_R + \bar \nu_R M S_L +\frac{1}{2}\bar S^C_L \mu S_L + H.C.,
\end{equation}
where $M_D = \frac{Y_D v_{\eta'}}{\sqrt{2}}$. These mass terms recover the characteristic mass matrix of the ISS mechanism given in  Eq. (\ref{mainmatrix}). Then all approach developed in Sec. II is applicable here. 

In what concern the contributions to the g-2  of the muon due to the heavy neutrinos $N$'s, they arise from the interactions of these neutrinos with the charged gauge bosons $W^{\pm}$ and $W^{\prime \pm}$,
\begin{eqnarray}
 {\cal L}^{CC}_{N,\mu, W,W'}&=&-\dfrac{g}{2\sqrt{2}}{\bar \mu } (1-\gamma_5)\gamma^\mu [({\cal M_D}^\dagger)[{\cal M_R}^\dagger]^{-1}U_R]_{2k}N_{k}W^-_\mu\\ \nonumber
 &-& \dfrac{g}{2\sqrt{2}}{\bar \mu } (1-\gamma_5)\gamma^\mu [({\cal M_D}^\dagger)[{\cal M_R}^\dagger]^{-1}U_R]_{2k}N_{k}W'^-_\mu + H.c.
 \label{CCM3312}
\end{eqnarray}
${\cal M_D}$, ${\cal M_R}$ and $U_R$ are exactly those in Sec. II. Our task here is to check if this new interactions give significant contributions to the g-2  of the muon. 

\subsection{3-3-1 model with right-handed neutrinos}
In this version of the 3-3-1 models\cite{PhysRevD.47.2918,PhysRevD.22.738,PhysRevD.50.R34} the third component of the leptonic triplet is occupied by the right-handed neutrinos invoked by the ISS mechanism\cite{PhysRevD.86.035007}

\begin{equation} f_{a_L} =
\begin{pmatrix}
   \nu_{a_L}  \\
    l_{a_L}  \\
    (\nu_{a_R})^c
    \end{pmatrix}.
\end{equation}
This version shares  the same main features that the minimal one in respect to anomaly cancellation, electric charged quantization and strong-CP problem with the additional one of having natural dark matter candidates\cite{de_S_Pires_2007,PhysRevD.83.065024,Dutra:2021lto}. However, it is a new model with its gauge sector being composed by the standard gauge bosons and five other ones as a new $Z^{\prime}$, two new single charged gauge bosons $W^{\prime \pm}$ and two neutral, but non-hermitian,  gauge bosons $U^{0}\,,\,U^{0 \dagger}$. The interactions of these gauge bosons with the neutral and charged currents is found in \cite{PhysRevD.53.437}. It was checked in  \cite{deJesus:2020ngn} that all these interactions are not enough to accommodate the recent muon g-2 result, and again this provides a strong motivation for we check if the implementation of the ISS mechanism would give significant contributions to the g-2 of the muon.

For the implementation of the mechanism, we just need to add three new neutral fermions in the singlet form, $S_L$. In this model the minimal set of scalars necessary to break spontaneously the symmetry and generate masses for all massive particles, except neutrinos, are three triplets of scalars $\eta'$, $\rho$, $\chi$.  With this minimal particle content we form the following terms that will trigger the ISS mechanism
\begin{equation}
    {\cal L}\supset    Y_{D_{ab}}\epsilon_{lmn}\overline{\left(f_{aL}\right)_{l}^c}\rho^*_m (f_{bL})_{n}+Y^{\prime}_{D_{ab}}\bar f_{aL} \chi (S_{bL})^C 
    +\frac{1}{2}\bar S^C_L \mu S_L + H.C..
\end{equation}
When $\rho$ and $\chi$ develop their vev's, $\langle \rho\rangle=\frac{v_{\rho}}{\sqrt{2}}$ and $\langle \chi \rangle=\frac{v_\chi}{\sqrt{2}}$, we have the following mass terms,
\begin{equation}
    {\cal L} = \bar \nu_L M_D \nu_R + \bar \nu_R M S_L +\frac{1}{2}\bar S^C_L \mu S_L + H.C.,
\end{equation}
where $M_D = \frac{Y_D v_\rho}{\sqrt{2}}$ and $M = \frac{Y^{\prime}_D v_\chi}{\sqrt{2}}$ . These mass terms recover the characteristic mass matrix of the ISS mechanism given in  Eq. (\ref{mainmatrix}). Then all approach developed in Sec. II is applicable here with specific care because, now, $M_D$ is an antisymmetric matrix and, due to this, we can not take it diagonal. But it is possible to take it degenerated. In this case we have
\begin{equation}
M_D=
\begin{pmatrix}
   0&m_D&m_D\\
   -m_D&0&m_D\\
   -m_D&-m_D&0
\end{pmatrix}.
\end{equation}
Following this, taking $M=m_N I$, we have for the nonunitary parameter, defined in Eq. (\ref{etaparameter})
\begin{equation}
\eta=\dfrac{m_D^2}{2 m_N^2}
\begin{pmatrix}
2 & 1& -1\\
1& 2& 1\\
  -1& 1& 2
\end{pmatrix}.
\end{equation}

In what concern the contributions to the g-2  of the muon due to the heavy neutrinos $N$'s, they arise from the interactions of these neutrinos with the charged gauge bosons $W^{\pm}$ and $W^{\prime \pm}$,
\begin{eqnarray}
 {\cal L}^{CC}_{N,\mu, W,W'}&=&-\dfrac{g}{2\sqrt{2}}{\bar \mu } (1-\gamma_5)\gamma^\mu [({\cal M_D}^\dagger)[{\cal M_R}^\dagger]^{-1}U_R]_{2k}N_{k}W^-_\mu\\ \nonumber
 &-& \dfrac{g}{2\sqrt{2}}{\bar \mu } (1-\gamma_5)\gamma^\mu [U_R]_{2k}N_{k}W'^-_\mu + H.c.,
 \label{CCM331}
\end{eqnarray}
with
\begin{equation}
{\cal M_D}{\cal M_R}^{-1}U_R=
\dfrac{m_D}{\sqrt{2} m_N}
\begin{pmatrix}
-1& -1& 0& -1& -1& 0\\
-1& 0& 1& -1& 0& 1\\
  0& 1& 1& 0& 1&1
  \end{pmatrix}.
\end{equation}

\begin{figure}[h!]
	\centering
	\includegraphics[width=8.5cm]{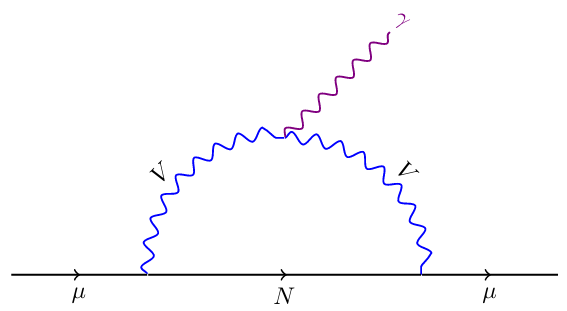}
	\caption{New Feynman diagram contributing to the muon g-2. $N$ are the  heavy pseudo-Dirac neutrinos characteristic of the ISS mechanism and $V^\pm=W^{\prime \pm}\,,\, W^{\pm}$.  }
	\label{fig:331iss}
\end{figure}

Observe that all the  new contributions to g-2  of the muon, in all three cases, are due to interactions involving charge currents and single charged gauge bosons as represented in Fig. \ref{fig:331iss} for $V$ being the standard gauge boson $W^{\pm}$ or  $W^{\prime \pm}$. For previous works addressing g-2 into this model, see Refs. \cite{Hue:2020wnn,Hue:2021xap}

We are now ready to calculate such contributions in all three cases. We do this and present our numerical results in the next section.

\section{Numerical results}

For our numerical calculations, we made use of the code  \cite{Mathematica}, developed in \cite{deJesus:2020ngn}. Then, after configuring this algorithm for each case, we solved numerically the integral in Eq. (\ref{aN}).

\begin{figure}[h]
\centering
     \includegraphics[scale=0.85]{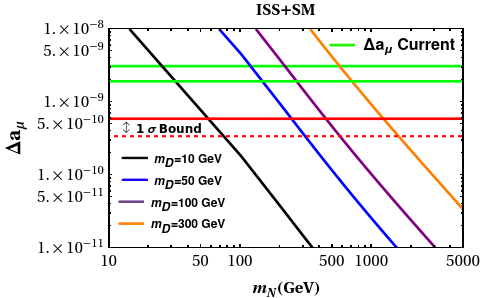} 
    \caption{ Overall contribution to $\Delta a_\mu$ from the minimal extension of the SM that accommodates ISS depending on different sterile neutrino masses $m_N$ and Dirac masses $m_D$. The green bands are delimited by  $\Delta a_\mu = (251 \pm 59) \times 10^{- 11}$. The current $1 \sigma$ bound is found by requiring $ \Delta a_\mu < 59 \times 10^{- 11}$ while the projected bound is obtained for $\Delta a_\mu < 34 \times 10^{- 11}$. We can see that, for values of $m_D$ from $10$ up to $300$ GeV, there are reasonable values for $m_N$  that explains $\Delta a_\mu$. }    
\label{fig:SM}
\end{figure}

\begin{figure}[h]
\centering
 
    \includegraphics[width=7.5cm]{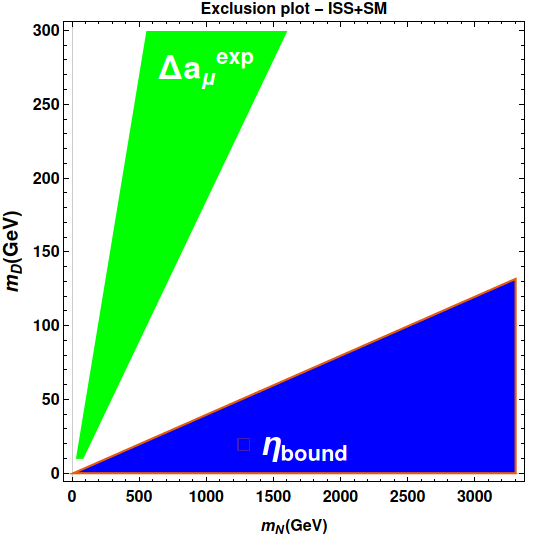} 
\caption{ Exclusion plot, where the green region is the parametric space that explains g-2 of the muon and the blue is the one that respects the present bound on the nonunitarity parameter $\eta_{\mbox{bound}}$.
}    
\label{fig:SM2}
\end{figure}

Our first result is presented in Figs. \ref{fig:SM} and \ref{fig:SM2}, where we consider the case of the ISS mechanism implemented into the Standard Model (ISS + SM). Notice that the new interactions in Eq. (\ref{ISSCCSM}) involving heavy neutrinos explain the recent muon g-2 results for $m_N$ varying from few tens up to thousands of GeV's.   However, the current bound on the nonunitarity parameter $\eta$,  Eq. (\ref{etaSM}),  requires the model to live in a different corner of the parameter space, delimited by the blue region. Thus, the regions do not overlap, and we conclude that one cannot simultaneously explain g-2 and be consistent with nonunitarity.  This means that the ISS mechanism, on its own, is not able to explain the muon g-2 result, at least when implemented in the standard model.    

\begin{figure}[h]
    \centering
    \includegraphics[scale=1.18]{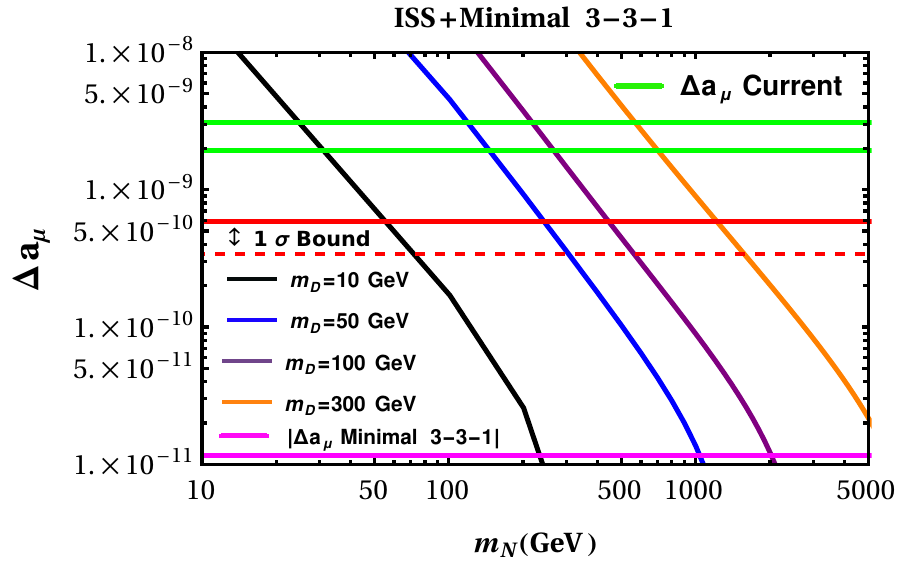} 
    \caption{ Overall contribution to $\Delta a_\mu$ from the minimal 3-3-1 model with ISS depending on different sterile neutrino masses $m_N$ and Dirac masses $m_D$. The pink line represents the absolute value of the total contribution for muon g-2 from minimal 3-3-1, fixing $m_{W^{\prime}}=5$ TeV and $m_{U^{++}}=5$ TeV. After the introduction of six neutrinos, the total contribution for $\Delta a_\mu$ can explain the muon g-2 for some values of $m_D$ and $m_N$.  
} \label{fig:minimal331}
\end{figure}

\begin{figure}[h]
    \centering
    \includegraphics[width=7.5cm]{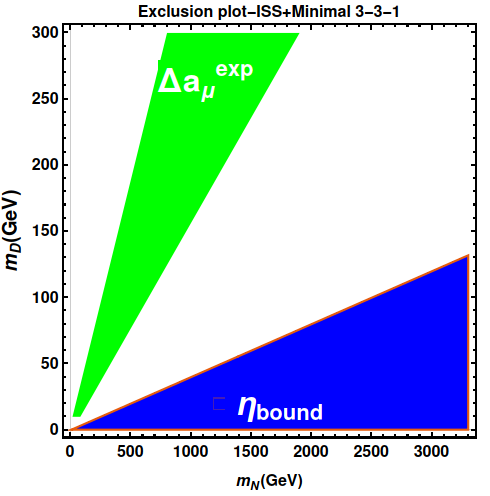} 
\caption{ Exclusion plot where the green region is the parametric space that explains g-2 of the muon and the blue is the one that respects the present bound on the nonunitarity parameter $\eta_{\mbox{bound}}$.
} \label{fig:minimal3312}
\end{figure}

In Figs. \ref{fig:minimal331} and \ref{fig:minimal3312}  we present our results for the case of the ISS mechanism implemented into the minimal 3-3-1 model. For minimal 3-3-1 model and for the 3-3-1 with right-handed neutrinos, we calculate the total contribution of $\Delta a_\mu$ for fixed gauge boson masses. These masses are set to be sufficiently high to avoid collider bounds, namely $m_{U^{++}}=5$ TeV   and $m_{W'}=5$ TeV \cite{deJesus:2020ngn} . We see, in the left-panel, that the minimal 3-3-1 model definitely cannot explain the g-2 result.  This happens mostly because the doubly-charged contribution to $\Delta a_\mu$ is dominant and negative. However, after extending the minimal 3-3-1 to accommodate the ISS, we observe new contributions for the muon g-2 in the same manner as the previous case. These new contributions are due to heavy neutrinos with interactions given in Eq. (\ref{CCM331}) and displayed in Fig. \ref{fig:331iss} with $V$ being due to the charged gauge bosons $W^{\pm}$ and $W^{\prime \pm}$. We present in Fig \ref{fig:minimal331} our results in the plane $m_D\times m_N$ .  It is already clear that the nonunitarity constraint prevents the ISS to explain g-2. We plot our result in a different plane in Fig \ref{fig:minimal3312} to clearly show that these datasets are mutually exclusive. We display the $\Delta a_\mu$ favored region delimited by green lines and the  $\eta_{\mbox{bound}}$ with a purple line towards the bottom of the figure. Note that the nonunitarity bound lies orders of magnitude below the favored region for g-2. In summary, ISS is not the solution to g-2. 

\begin{figure}[h]
    \centering
    \includegraphics[scale=0.52]{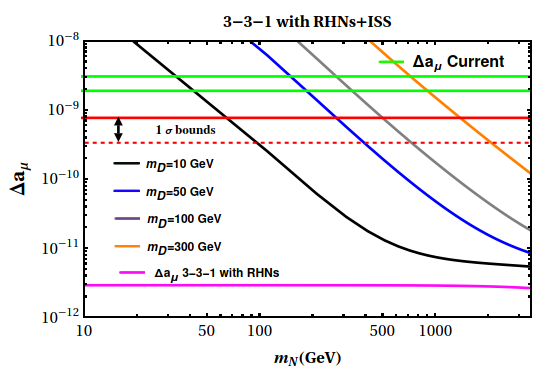} 
 \caption{ Overall contribution to $\Delta a_\mu$ from the 3-3-1 model with right-handed neutrino (RHN)  with ISS depending on different sterile neutrino masses $m_N$ and Dirac masses $m_D$. The pink line represents the absolute value of the total contribution for muon g-2 from 3-3-1 with RHNs. After the introduction of three neutrinos, the total contribution for $\Delta a_\mu$ may explain the muon g-2 for some set of values for $m_D$ and $m_N$.
}  \label{fig:331RH}
\end{figure}

\begin{figure}[h]
    \centering

    \includegraphics[width=7.5cm]{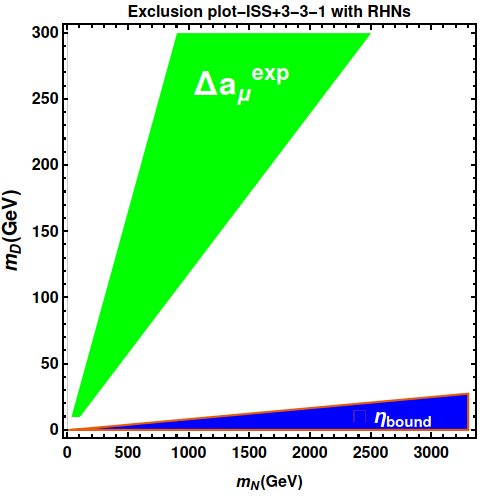} 
 \caption{Exclusion plot where the green region is the parameter space that explains g-2 of the muon and the blue is the one that respects the present bound on the nonunitarity parameter $\eta_{\mbox{bound}}$.
}  \label{fig:331RH2}
\end{figure}

Finally, we show in Figs. \ref{fig:331RH} and \ref{fig:331RH2} our results for the case of the ISS implemented into the 3-3-1 model with right-handed neutrinos. Looking at the numerical results, we arrive at the same conclusion.  The ISS mechanism cannot address g-2 even in the scope of a 3-3-1 model with right-handed neutrinos.

\clearpage

\clearpage

\section{Discussion}

We would like to emphasize that our conclusion concerning the inverse seesaw is applicable to the minimal version of the inverse seesaw, in other words, when it is implemented in the standard model, and to the 3-3-1 models studies. Our conclusion is not valid to all possible implementations of the inverse seesaw as one can always add fields in a way to change the contribution to  the g-2 or circumvent the non-unitarity bounds, and consequently find different results. Obviously, such constructions are necessarily more complex. A concrete example can be found in \cite{Khalil:2015wua,Cao:2019evo,Cao:2021lmj}.

\section{Conclusions}
We revisited the inverse seesaw mechanism and assessed the possibility to explain g-2 when nonunitarity constraints are considered. Firstly, we added the necessary fields to the Standard Model spectrum to realize inverse seesaw mechanism. Later, we implemented the inverse seesaw in the minimal 3-3-1 model, and in the 3-3-1 model with right-handed neutrinos. We studied the impact of the new neutral singlet fields on the lagrangians relevant for g-2 and nonunitarity studies. In all three cases, we conclusively showed that the ISS mechanism cannot address g-2 due to the stringent nonunitarity constraint that forces the masses of the particles to be orders of magnitude higher than the one required to explain g-2. This conclusion is clear in Fig. \ref{fig:331RH}. In summary, our findings show that the nonunitarity constraint is severe enough to avoid the inverse seesaw mechanism to explain the recent muon g-2 result within these frameworks.

\section{Acknowledgments}

J.P.P has received funding/support from the European Union’s Horizon 2020 research and innovation programme under the Marie Skłodowska-Curie grant agreement No 860881-HIDDeN. C.A.S.P  was supported by the CNPq research grants No. 304423/2017-3. Y.S.V acknowledges the financial support from CAPES under Grant No. 88882.375870/2019-01. FSQ is supported by the S\~ao Paulo Research Foundation (FAPESP) through grant 2015/158971, ICTP-SAIFR
FAPESP grant 2016/01343-7, CNPq grants 303817/2018-6 and 421952/2018 – 0, and the Serrapilheira Institute (grant
number Serra - 1912 – 31613). We dedicate this work to the memory of Joel Batista da Fonseca Neto, a great professor from UFPB who passed away recently.  This work was supported by a grant from the Simons Foundation Award Number 884966, AF.

\end{document}